\documentclass[prb,aps,showpacs,onecolumn]{revtex4}
\usepackage{amsfonts}
\usepackage{graphicx}
\usepackage{amsmath}
\usepackage{bm}

\input{tcilatex}
\begin{document}

\title{On Peres approach to Fradkin-Bacry-Ruegg-Souriau's perihelion vector}
\author{Y. Grandati, A. B\'{e}rard and and H. Mohrbach }
\affiliation{Laboratoire de Physique Mol\'eculaire et des Collisions, ICPMB, IF CNRS
2843, Universit\'e Paul Verlaine, Institut de Physique, Bd Arago, 57078
Metz, Cedex 3, France}

\begin{abstract}
We solve explicitely the differential system obtained by Peres for the
construction of a conserved vector associated to any central potential. We
then obtain a very direct access to the discontinuous behavior of this
Fradkin-Bacry-Ruegg-Souriau perihelion vector.
\end{abstract}

\maketitle

%\date{\today}

.

\section{Introduction}

Since the pioneer works of Ermanno, Bernouilli and Laplace during eighteenth
century\cite{leach,goldstein,whittaker} we know that the Kepler system
presents a specific vector conserved quantity, the so-called
Laplace-Runge-Lenz vector, which provides in particular a very simple access
to the orbit equation. Today, the existence of generalized
Laplace-Runge-Lenz vectors for various extensions of the Kepler problem have
been put in evidence\cite{leach}.

This\ supplementary constant of motion has a dynamical origin which is
directly linked to the existence, for these systems, of a symmetry group
larger than the space-time's one\cite{bander,greenberg,mac,cordani}. As
established by Jauch and Hill\cite{jauch} in the 2D case and by Fradkin\cite%
{fradkin} in the 3D case, the isotropic harmonic oscillator shares the same
feature and admits also a supplementary dynamical conserved quantity but of
tensorial type, the Fradkin-Jauch-Hill tensor\cite%
{greenberg,leach,sivardiere}.

Bacry, Ruegg, Souriau\cite{bacry} and Fradkin\cite{fradkin2} made an
important additional step when they showed that all three dimensional
dynamical problems involving central potentials possess the extended
symmetry algebra O$_{4}$ and SU$_{3}$, a result soon generalized by Mukunda%
\cite{mukunda} for any hamiltonian system with n degrees of freedom. As
noted by Bacry, Ruegg and Souriau in their seminal paper\cite{bacry}, the
general ability to construct such a local Lie algebra of dynamical symmetry
is not surprising since all the 2n-dimensional symplectic manifolds are
locally isomorphic\cite{souriau}. However the following step to ensure a
global dynamical symmetry is to verify that the finite canonical
transformations generated by this algebra form a group, a feature of the
Kepler problem. As established by Stehle and Han\cite{stehle}, the existence
of such a \textquotedblright global\textquotedblright\ higher symmetry
group, admitting finitely multivalued realizations on the phase space of the
system, is a necessary condition for the existence of degeneracies.

Bacry, Ruegg, Souriau\cite{bacry}, as much as Fradkin\cite{fradkin2}, gave
an explicit construction scheme for a vector with constancy properties
associated to every central potential, Fradkin's derivation being very
direct and explicit. Bacry, Ruegg and Souriau already noted that such a
vector is exceptionally one-valued, as in the Kepler case. The first
explicit demonstration of the multivalued behavior of these
Fradkin-Bacry-Ruegg-Souriau (FBRS) vectors have been established by
Serebrennikov, Shabad\cite{sereb}, Buch and Denman\cite{buch} when these
authors pointed out that in specific cases Fradkin vectors are in fact only
piecewise conserved.

In 1979, Peres\cite{peres}, using a distinct approach, rediscovered the FBRS
generalized Laplace-Runge-Lenz vector. He obtained a differential system for
the coefficients of the vectors which presents singularities at the apsidal
positions. Without solving explicitely the differential system, Peres
infered from this singular behavior the existence of discontinuous changes
of direction of the vector which is then only piecewise constant. The
correspondence between Fradkin's and Peres approaches have been studied by
Yoshida\cite{yoshida}.

A complete answer to the question concerning the status of conserved
quantity which could be attributed to the FBRS vector has been given by
Holas and March\cite{holas}. They showed explicitely that at each pericenter
(or apocenter, depending on the chosen initial conditions), the FBRS vector
changes its direction abruptly from an angle double of the apsidal one. A
last attempt to obtain a "true" vector constant of motion in every
conservative central force field has been made by Yan\cite{yan} but, as
shown a short time after by Holas and March\cite{holas2}, Yan's construction
coincides in fact with the FBRS perihelion vector and then presents the same
discontinuities. The generalized Laplace-Runge-Lenz vector can also be
deduce in a direct and elegant manner from the equation of motion\cite{leach}%
.

Note finally that, in a very recent contribution, Ballesteros and al.\cite%
{ball} use FBRS type vectors to establish an optimal extension of Bertrand's
theorem\cite{bert,grandati} to curved spaces.

In this paper, we give an explicit solution to the differential system
obtained by Peres\cite{peres}. Using then a complex formulation,
particularly convenient for the treatment of planar problems\cite%
{grandati1,grandati2}, we recover in a very simple manner the discontinuous
behavior of the FBRS vector studied by Holas and March\cite{holas}.

\section{Basics of 2D motion in a central potential}

\subsubsection{Complex formulation}

We consider a planar motion $\overrightarrow{r}\left( t\right) =\binom{x(t)}{%
y(t)}_{(O,\overrightarrow{u_{x}},\overrightarrow{u_{y}})}$ for a particle of
mass $m=1$ submitted to a potential $U(\overrightarrow{r})$, eventually
singular at the origin. We choose to adopt a complex formulation where we
represent the position by its corresponding affix $z(t)=x(t)+iy(t)$, the
potential being then viewed as a real valued function of $z$, $U(z,\overline{%
z})$, defined on $\mathbb{C}$ or $\mathbb{C}^{\ast }$. The gradient of any
real valued function of $z$ and $\overline{z}$ is then given by : 
\begin{equation}
\quad \overrightarrow{\nabla }U(\overrightarrow{r})\rightarrow 2\frac{%
\partial U(z,\overline{z})}{\partial \overline{z}}
\end{equation}

where $\frac{\partial }{\partial \overline{z}}=\overline{\frac{\partial }{%
\partial z}}=\frac{1}{2}\left( \frac{\partial }{\partial x}+i\frac{\partial 
}{\partial y}\right) $.

The equation of motion for our system takes the form:

\begin{equation}
\overset{..}{z}+2\frac{\partial U(z,\overline{z})}{\partial \overline{z}}=0
\label{eqmouv1}
\end{equation}

where the dot represents the time derivative.

For two vectors $\overrightarrow{A}$ and $\overrightarrow{B}$ in the $(O,%
\overrightarrow{u_{x}},\overrightarrow{u_{y}})$ plane, $\overrightarrow{A}%
\times \overrightarrow{B}$ is represented in complex form by the real
quantity $\func{Im}\left( \overline{A}B\right) $, $A$ and $B$ be being the
complex affixes of $\overrightarrow{A}$ and $\overrightarrow{B}$
respectively.

The angular momentum (which necesseraly conserves its direction orthogonal
to the plane) $\overrightarrow{L}\left( t\right) =\overrightarrow{r}\left(
t\right) \times \overset{.}{\overrightarrow{r}}\left( t\right) =L\left(
t\right) \overrightarrow{k}$ , admits then the following correspondent :

\begin{equation}
L=\func{Im}\left( \overline{z}(t)\overset{.}{z}(t)\right) =\frac{1}{2i}%
\left( \overline{z}\overset{.}{z}-\overset{.}{\overline{z}}z\right)
\label{momcin1}
\end{equation}

In the case of a central potential $U(z,\overline{z})=U(\left\vert
z\right\vert )=U(r)$, we have:

\begin{equation}
\quad \overrightarrow{\nabla }U(r)\rightarrow \frac{z}{r}U^{\prime }(r)
\end{equation}

where $f^{\prime }(x)=\frac{df(x)}{dx}$ is the usual derivative of a
function $f$ of one variable $x$.

Eq.(\ref{eqmouv1}) becomes simply:

\begin{equation}
\overset{..}{z}+\frac{z}{r}U^{\prime }(r)=0  \label{eqmouv2}
\end{equation}

Using this result, the angular momentum's conservation is immediate (see Eq.(%
\ref{momcin1})):

\begin{equation}
\overset{.}{L}=\frac{1}{2i}\left( \overline{z}\overset{..}{z}-\overset{..}{%
\overline{z}}z\right) =0
\end{equation}

\subsubsection{Radial equation of motion}

If we use a polar representation for $z$, $z=re^{i\theta }$, with $\theta
\in \left[ 0,2\pi \right[ $, we have:

\begin{equation}
\overset{.}{\widehat{\left( r^{2}\right) }}=2\overset{.}{r}r=\overset{.}{z}%
\overline{z}+z\overset{.}{\overline{z}}
\end{equation}

which gives:

\begin{equation}
\overset{.}{r}=\frac{1}{2r}\left( \overset{.}{z}\overline{z}+z\overset{.}{%
\overline{z}}\right) =\frac{1}{r}\left( z\overset{.}{\overline{z}}+iL\right)
\end{equation}

and (since $e^{2i\theta }=\frac{z}{\overline{z}}$) :

\begin{equation}
\overset{.}{\theta }(t)=\frac{\overline{z}\overset{.}{z}-\overset{.}{%
\overline{z}}z}{2ir^{2}}=\frac{L}{r^{2}(t)}  \label{thetapoint}
\end{equation}

Then, we can write:

\begin{equation}
\overset{.}{z}=\overset{.}{r}e^{i\theta }+i\overset{.}{\theta }re^{i\theta
}=\left( \overset{.}{r}+i\frac{L}{r}\right) \frac{z}{r}  \label{vitesse}
\end{equation}

and:

\begin{equation}
\overset{..}{z}=\left( \frac{\overset{..}{r}}{r}-\frac{L^{2}}{r^{4}}\right) z
\end{equation}

Then using Eq.(\ref{eqmouv2}), we deduce an equation for the the radial
motion:

\begin{equation}
\overset{..}{r}(t)=\frac{L^{2}}{r^{3}(t)}-U^{\prime }(r(t))
\label{eqradiale}
\end{equation}

which is readily integrated as :

\begin{equation}
\left( \overset{.}{r}\right) ^{2}=2\left( E-\frac{L^{2}}{2r^{2}}-U(r)\right)
\end{equation}

$E$ is a constant of integration, which can be identified with the energy of
the system, since from Eq.(\ref{vitesse}) we have:

\begin{equation}
\frac{1}{2}\left| \overset{.}{z}\right| ^{2}=\frac{1}{2}\left( \frac{\overset%
{.}{r}}{r}+i\frac{L}{r^{2}}\right) z\left( \frac{\overset{.}{r}}{r}-i\frac{L%
}{r^{2}}\right) \overline{z}=\frac{1}{2}\left( \left( \overset{.}{r}\right)
^{2}+\frac{L}{r^{2}}\right) =E-U(r)
\end{equation}

If we note $T(r)=E-U(r)$, Eq.(\ref{eqradiale}) takes the form:

\begin{equation}
\left( \overset{.}{r}(t)\right) ^{2}=\frac{L^{2}}{r^{2}}f(r(t)),
\label{eqfond1}
\end{equation}

where :

\begin{equation}
f(r)=\frac{2r^{2}}{L^{2}}T(r)-1=\frac{2r^{2}}{L^{2}}(E-V_{L}(r)),
\label{deff}
\end{equation}

$V_{L}(r)=U(r)+\frac{L^{2}}{2r^{2}}$ being the radial effective potential.

From Eq.(\ref{eqfond1}) and Eq.(\ref{deff}), we see that the radial
oscillation motion presents, for bounded orbits, two turning point $r_{m}$
and $r_{M}$, corresponding to the pericenters et apocenters (apsidal
positions) of the orbital motion. They are roots of the equation:

\begin{equation}
f(r)=0\Leftrightarrow V_{L}(r)=E  \label{point de rebr}
\end{equation}

In the sequel, we will suppose that $r_{m}$ and $r_{M}$, are simple roots,
that is $V_{L}^{\prime }(r_{m})<0$ and $V_{L}^{\prime }(r_{M})<0$.

At the apsidal positions, $\overset{.}{r_{m}}=\overset{.}{r_{M}}=0$ and:

\begin{equation}
\overset{.}{z}_{j}=i\overset{.}{\theta _{j}}z_{j}=i\frac{L}{r_{j}^{2}}%
z_{j},\quad j=m,M  \label{vitaps}
\end{equation}

The $i$ factor in the right hand side of Eq.(\ref{vitaps}) implies that, for
apsidal positions, the instantaneous speed vector is orthogonal to the
position vector.

Note that if $z(t)$ and $\overset{.}{z}(t)$ are continuous and
differentiable for every values of $t$, the differentiability domain of $%
r(t) $ is stricly limited to the real interval $\left] r_{m},r_{M}\right[ $.

To extract from Eq.(\ref{eqfond1}) the radial celerity $\overset{.}{r}(t)$
as a real-valued function, the square root has to be constructed with a
pre-defined sign. To take account of the turn back in the radial motion,
this sign changes when $r$ reaches the extremal values $r_{m}$ and $r_{M}$,
which corresponds to change the square root determination.

We obtain:

\begin{equation}
\overset{.}{r}(t)=\pm \frac{L}{r(t)}\sqrt{f(r(t))}  \label{eqfond2}
\end{equation}

or:

\begin{equation}
dr(t)=\pm \frac{L}{r}\sqrt{f(r)}dt
\end{equation}

The + sign is chosen when $r$ moves from $r_{m}$ to $r_{M}$ (increasing
phase of the radial oscillation, $dr(t)>0$) and the - sign during the
reverse motion (decreasing phase of the radial oscillation, $dr(t)<0$). To
be more precise, we will index each phase of the motion by an integer number 
$k\in \mathbb{N}$, the even value $k=2n$ corresponding to the decreasing
phase of the $n$-th period (except for $n=0$) and the odd value $k=2n+1$
corresponding to the increasing phase of the same period. With this
convention, we then have:

\begin{equation}
dr(t)=(-1)^{k+1}\frac{L}{r(t)}\sqrt{f(r(t))}dt  \label{drt}
\end{equation}

and:

\begin{equation}
\overset{.}{r}(t)=(-1)^{k+1}\frac{L}{r(t)}\sqrt{f(r(t))}  \label{eqfond3}
\end{equation}

Eq.(\ref{drt}) gives implicitely the solution of the radial equation of
motion Eq.(\ref{eqradiale}).

\subsubsection{Orbital equation and $r$-parametrization of the motion}

In the $k$-phase of the motion ,in every point except the apsidal ones, $%
\overset{.}{r}(t)$ is a monotonic function (see Eq.(\ref{eqfond3})) and we
can choose to parametrize the motion by $r$ rather than by $t$.

For instance:%
\begin{equation}
\frac{d\theta }{dr(t)}=\frac{\overset{.}{\theta }}{\overset{.}{r}}=(-1)^{k+1}%
\frac{1}{r\sqrt{f(r)}}  \label{dtheta}
\end{equation}

where we have used Eq.(\ref{eqfond3}) and Eq.(\ref{thetapoint}).

Choosing a reference time $t_{0}$ during the same phase, we can write:

\begin{equation}
\theta \left( r(t)\right) =\theta \left( r_{0}\right)
+(-1)^{k+1}\int_{r_{0}}^{r(t)}\frac{d\rho }{\rho \sqrt{f(\rho )}}
\end{equation}

with $r(t_{0})=r_{0}$ and the orbital equation for this phase writes:

\begin{equation}
\theta \left( r\right) =\theta \left( r_{0}\right)
+(-1)^{k+1}\int_{r_{0}}^{r}\frac{d\rho }{\rho \sqrt{f(\rho )}}  \label{eqorb}
\end{equation}

We have to be more careful if we want to use $r$ as a global parameter for
the motion. Indeed, since $r(t)$ is a periodic function of $t$, $t(r)$\ and
every monotonic function of $t$, as $\theta $\ (see Eq.(\ref{thetapoint})),
will be multivalued. Therefore, at every $k$ phase of the\ motion will be
attached a different branch of the function.

\bigskip For simplicity, we choose the initial condition:

\begin{equation}
\left\{ 
\begin{array}{c}
r(t=0)=r_{m} \\ 
\theta \left( t=0\right) =0%
\end{array}%
\right.
\end{equation}

Then, the different determinations of $\theta \left( r\right) $
corresponding to each phase of the motion are given by:%
\begin{eqnarray}
\text{In the first phase of the motion }\text{: } &&\theta \left(
r(t)\right) =g(r(t)) \\
\text{In the second phase of the motion }\text{: } &&\theta \left(
r(t)\right) =\Phi -\int_{r_{M}}^{r(t)}\frac{d\rho }{\rho \sqrt{f(\rho )}}%
=2\Phi -g(r(t))  \notag \\
\text{In the third phase of the motion }\text{: } &&\theta \left(
r(t)\right) =2\Phi +g(r(t))  \notag \\
&&...
\end{eqnarray}

where: 
\begin{equation}
g(r(t))=\int_{r_{m}}^{r(t)}\frac{d\rho }{\rho \sqrt{f(\rho )}},  \label{g}
\end{equation}

$\Phi =g(r_{M})=\int_{r_{m}}^{r_{M}}\frac{dr^{\prime }}{r^{\prime }\sqrt{%
f(r^{\prime })}}$ being the apsidal angle.

More generally, we will write:

\begin{equation}
\text{In the }k\text{-th phase of the motion }\text{: }\theta _{k}\left(
r(t)\right) =2n\Phi +(-1)^{k+1}g\left( r(t)\right)  \label{teta}
\end{equation}

where $n$ is the integral part of $\frac{k}{2}$.

\bigskip From Eq.(\ref{eqfond3}) and Eq.(\ref{dtheta}),\ we deduce the
following expression for the instantaneous velocity in the $k$-th phase:

\begin{equation}
\overset{.}{z}(t)=\left( \left( -1\right) ^{k+1}\sqrt{f(r(t))}+i\right) 
\frac{L}{r^{2}(t)}z(t)=\frac{L}{r(t)}\left( \left( -1\right) ^{k+1}\sqrt{%
f(r(t))}+i\right) e^{i\theta _{k}\left( r(t)\right) }  \label{zpointder}
\end{equation}

$\theta _{k}\left( r\right) $ being given by Eq.(\ref{teta}).

Note that, since $\overset{.}{z}=\lambda z$ where $\lambda $ has a nonzero
imaginary part, Eq.(\ref{zpointder}) ensures the non-colinearity of $%
\overrightarrow{r}$ and $\overset{.}{\overrightarrow{r}}$.

In the sequel, we will have to use the vector $\overset{.}{\overrightarrow{r}%
}\times \overrightarrow{L}$which has the following complex correspondent :

\begin{equation}
\overset{.}{\overrightarrow{r}}(t)\times \overrightarrow{L}\rightarrow iL%
\overset{.}{z}(t)=-\left( 1+\left( -1\right) ^{k}i\sqrt{f(r(t))}\right) 
\frac{L^{2}}{r^{2}(t)}z(t)  \label{Lvectrpoint}
\end{equation}

We see immediately that $\overrightarrow{L}\times \overset{.}{%
\overrightarrow{r}}$ and $\overrightarrow{r}$ are always linearly
independent, except at the apsidal positions.

\section{Peres approach to the FBRS vector}

In the special case of the Kepler system $U(r)=-r^{-1}$ , the system admits
the Laplace-Runge-Lenz vector, $\overrightarrow{\mathcal{A}}_{LRL}$ $=%
\overset{.}{\overrightarrow{r}}\times \overrightarrow{L}-\frac{1}{r}%
\overrightarrow{r}$, as a supplementary conserved quantity\cite{leach}. In
the complex formulation used above, this gives :

\begin{equation}
\overrightarrow{\mathcal{A}}_{LRL}\rightarrow \mathcal{A}_{LRL}=iL\overset{.}%
{z}-\frac{1}{r}z  \label{LRL}
\end{equation}

Following Peres\cite{peres}, for the general central potential $U(r)$, we
will then look for a supplementary conserved vector of the form:

\begin{equation}
\overrightarrow{\mathcal{A}}=\overset{.}{\overrightarrow{r}}\times 
\overrightarrow{L}\left( \frac{r^{2}}{L^{2}}a(r)\right) +b(r)\overrightarrow{%
r}
\end{equation}

whose associated complex correspondent is:

\begin{equation}
\mathcal{A}=\left( \frac{r^{2}}{L^{2}}a(r)\right) iL\overset{.}{z}+b(r)z,
\label{defA}
\end{equation}

$a(r)$ and $b(r)$\ being two real valued functions that we will have to
determine (the factor $\frac{r^{2}}{L^{2}}$ in the first term has been
introduced for future convenience).

Using Eq.(\ref{Lvectrpoint}), we can rewrite $\mathcal{A}$ in the $k$-th
phase, as:

\begin{equation}
\mathcal{A}_{k}=\left( b(r)-a(r)+\left( -1\right) ^{k+1}i\sqrt{f(r)}%
a(r)\right) z  \label{fradkinvect}
\end{equation}

\subsubsection{Differential system for the coefficients of the FBRS vector}

On every time interval $I$ on which $\mathcal{A}$ is constant we must have:

\begin{equation}
\overset{.}{\mathcal{A}}=0
\end{equation}

that is, with Eq.(\ref{zpointder}):

\begin{equation}
0=\frac{d}{dt}\left( b(r)-a(r)+\left( -1\right) ^{k+1}i\sqrt{f(r)}%
a(r)\right) z+\left( b(r)-a(r)+\left( -1\right) ^{k+1}i\sqrt{f(r)}%
a(r)\right) \overset{.}{z}
\end{equation}

\begin{eqnarray}
&&\overset{.}{r}\left( b^{\prime }(r)-a^{\prime }(r)+\left( -1\right)
^{k+1}i\left( \sqrt{f(r)}a^{\prime }(r)+\frac{a(r)f^{\prime }(r)}{2\sqrt{f(r)%
}}\right) \right) z  \notag \\
&&+\left( b(r)-a(r)+\left( -1\right) ^{k+1}i\sqrt{f(r)}a(r)\right) \left(
\left( -1\right) ^{k+1}\sqrt{f(r)}+i\right) \frac{L}{r^{2}}z=0
\end{eqnarray}

Using Eq.(\ref{eqfond3}), this becomes:

\begin{equation}
(-1)^{k+1}\sqrt{f(r)}\left( b^{\prime }(r)-a^{\prime }(r)+\frac{b(r)-2a(r)}{r%
}\right) +i\left( \sqrt{f(r)}a^{\prime }(r)+\left( \frac{f^{\prime }(r)}{2%
\sqrt{f(r)}}+\frac{f(r)-1}{r}\right) a(r)+b(r)\right) =0
\end{equation}

Consequently, in the $k$-phase of the motion, on the interval of radial
values $\left] r_{m},r_{M}\right[ $, we must have:

\begin{equation}
\left\{ 
\begin{array}{c}
a^{\prime }(r)-b^{\prime }(r)+\frac{2a(r)-b(r)}{r}=0 \\ 
rf(r)a^{\prime }(r)+a(r)\left( r\frac{f^{\prime }(r)}{2}+f(r)-1\right)
+b(r)=0%
\end{array}%
\right.  \label{systemediff1}
\end{equation}

Extracting $a^{\prime }(r)$ from the second equation above and substituting
in the first one, we obtain, in matrix form:

\begin{equation}
\binom{a(r)}{b(r)}^{\prime }=\frac{-1}{rf(r)}\left( 
\begin{array}{cc}
r\frac{f^{\prime }(r)}{2}+f(r)-1 & 1 \\ 
r\frac{f^{\prime }(r)}{2}-f(r)-1 & f(r)+1%
\end{array}%
\right) \binom{a(r)}{b(r)}\text{ }  \label{systemediff2}
\end{equation}

\subsubsection{Exact solution}

Peres\cite{peres}, starting from a first order differential system
equivalent to Eq.(\ref{systemediff1}) transforms it in a second order
differential equation for the coefficient $a(r)$. Nevertheless, the
structure of this last equation is, at first sight, rather complicated and
the author restricts its analysis to a characterization of the singularities.

Things are much more transparent if we choose to work with a slightly
different unknown function.

Indeed, it is readily seen that the first equation in Eq.(\ref{systemediff1}%
) can be rewritten:

\begin{equation}
ra^{\prime }(r)+2a(r)=\left( rb(r)\right) ^{\prime }  \label{eq1}
\end{equation}

Introducing an auxiliary function $u(r)$ defined by:

\begin{equation}
a(r)=u^{\prime }(r),  \label{defdeu}
\end{equation}

the above equation Eq.(\ref{systemediff1}) becomes:

\begin{equation}
ru^{\prime \prime }(r)+2u^{\prime }(r)=\left( ru(r)\right) ^{\prime \prime
}=\left( rb(r)\right) ^{\prime }
\end{equation}

Then:

\begin{equation}
b(r)=\frac{1}{r}\left( ru(r)\right) ^{\prime }+b_{0}=u^{\prime }(r)+\frac{1}{%
r}u(r)  \label{defdeu2}
\end{equation}

where the integration constant $b_{0}$ has been chosen equal to $0$.

We can report the expressions Eq.(\ref{defdeu}) and Eq.(\ref{defdeu2}) of $%
a(r)$ and $b(r)$ in terms of $u(r)$ in the second equation of Eq.(\ref%
{systemediff1}). This gives:

\begin{equation}
\left( u^{\prime }(r)+\frac{1}{r}u(r)\right) ^{\prime }=\frac{-1}{rf(r)}%
\left( u^{\prime }(r)\left( r\frac{f^{\prime }(r)}{2}-f(r)-1\right)
+(u^{\prime }(r)+\frac{1}{r}u(r))\left( f(r)+1\right) \right)
\end{equation}

or :

\begin{equation}
u^{\prime \prime }(r)+\left( \frac{1}{r}+\frac{f^{\prime }(r)}{2f(r)}\right)
u^{\prime }(r)+\frac{1}{r^{2}f(r)}u(r)=0
\end{equation}

In a more compact form, we obtain the following second order differential
equation for $u(r)$:

\begin{equation}
u^{\prime \prime }(r)-\left( \log \left( \frac{1}{r\sqrt{f(r)}}\right)
\right) ^{\prime }u^{\prime }(r)+\left( \frac{1}{r\sqrt{f(r)}}\right)
^{2}u(r)=0  \label{eqpouru}
\end{equation}

or (see Eq.(\ref{g})):

\begin{equation}
u^{\prime \prime }(r)-\frac{g^{\prime \prime }\left( r\right) }{g^{\prime
}\left( r\right) }u^{\prime }(r)+\left( g^{\prime }(r)\right) ^{2}u(r)=0
\label{eqpouru2}
\end{equation}

The coefficients in Eq.(\ref{eqpouru2}) depending only on $g\left( r\right) $%
, if we define:

\begin{equation}
u(r)=v(g\left( r\right) )  \label{defdev}
\end{equation}

Eq.(\ref{eqpouru2}) becomes then:

\begin{equation}
\left( g^{\prime }\left( r\right) \right) ^{2}v^{\left( 2\right) }(g\left(
r\right) )+g^{\prime \prime }\left( r\right) v^{\left( 1\right) }(g\left(
r\right) )-\frac{g^{\prime \prime }\left( r\right) }{g^{\prime }\left(
r\right) }\left( g^{\prime }\left( r\right) v^{\left( 1\right) }(g\left(
r\right) )\right) +\left( g^{\prime }\left( r\right) \right) ^{2}v(g\left(
r\right) )=0
\end{equation}

that is, simply:

\begin{equation}
v^{\left( 2\right) }(g\left( r\right) )+v(g\left( r\right) )=0
\label{eqpourv}
\end{equation}

where $v^{\left( n\right) }(x)=\frac{d^{n}v}{dx^{n}}$.

The resolution of Eq.(\ref{eqpourv}) is immediate and gives:

\begin{equation}
u(r)=\alpha \cos g\left( r\right) +\beta \sin g\left( r\right) ,\quad \alpha
,\beta \in \mathbb{R}  \label{u(r)}
\end{equation}

Reporting this result in Eq.(\ref{defdeu}) and Eq.(\ref{defdeu2}), we
finally have:

\begin{equation}
\left\{ 
\begin{array}{c}
a(r)=\frac{1}{r\sqrt{f(r)}}\left( \beta \cos g\left( r\right) -\alpha \sin
g\left( r\right) \right) \\ 
b(r)=a(r)+\frac{1}{r}\left( \alpha \cos g\left( r\right) +\beta \sin g\left(
r\right) \right)%
\end{array}%
\right. ,\quad \alpha ,\beta \in \mathbb{R}  \label{sol1}
\end{equation}

that is:

\begin{equation}
\left\{ 
\begin{array}{c}
a(r)=\frac{\gamma }{r\sqrt{f(r)}}\cos \left( g\left( r\right) +\phi \right)
\\ 
b(r)=a(r)+\frac{\gamma }{r}\sin \left( g\left( r\right) +\phi \right)%
\end{array}%
\right. ,\quad \gamma \in \mathbb{R},\ \phi \in \left[ 0,2\pi \right[
\label{sol2}
\end{equation}

\subsubsection{Discontinuous behavior of $\mathcal{A}$}

The preceding result Eq.(\ref{sol2}) determines completely the form of $%
\mathcal{A}$ on $\left] r_{m},r_{M}\right[ $ during a given phase of the
motion. Indeed, inserting Eq.(\ref{sol2}) in Eq.(\ref{fradkinvect}), we
obtain:

\begin{equation}
\mathcal{A}_{k}=\frac{\gamma }{r}\left( \sin \left( g\left( r\right) +\phi
\right) +(-1)^{k+1}i\cos \left( g\left( r\right) +\phi \right) \right)
z(r)=(-1)^{k+1}i\gamma e^{(-1)^{k}i\left( g\left( r\right) +\phi \right)
}e^{i\theta _{k}\left( r\right) }
\end{equation}

Using Eq.(\ref{teta}), this becomes:

\begin{equation}
\mathcal{A}_{k}=(-1)^{k+1}i\gamma e^{(-1)^{k}i\left( g\left( r\right) +\phi
\right) }e^{i\left( 2n\Phi +(-1)^{k+1}g\left( r\right) \right)
}=(-1)^{k+1}\gamma ie^{(-1)^{k}i\phi +2ni\Phi }  \label{fadkinvect2}
\end{equation}

where $n$ is the integer part of $\frac{k}{2}$.

Under a more detailed form, we have:

\begin{equation}
\left\{ 
\begin{array}{c}
\mathcal{A}_{2n}=-\gamma e^{i\frac{\pi }{2}+i\phi +2ni\Phi } \\ 
\mathcal{A}_{2n+1}=\gamma e^{i\frac{\pi }{2}-i\phi +2ni\Phi }%
\end{array}%
\right.  \label{fradkinvect3}
\end{equation}

The integration constants $\gamma $ and $\phi $ for and determine then
respectively the modulus and argument of the $\overrightarrow{\mathcal{A}}$
vector affix. The freedom in the choice of the parameters $\gamma $ and $%
\phi $ induces that $\overrightarrow{\mathcal{A}}$ can be identified with
any vector of the plane. We have the same type of result for the specific
Kepler problem. Indeed, in this case, we can add to the Laplace-Runge-Lenz
vector $\overrightarrow{\mathcal{A}}_{K}$ a second conserved vector, $%
\overrightarrow{\mathcal{S}}_{K}=\overrightarrow{L}\times \overrightarrow{%
\mathcal{A}}_{K}$, the so-called Hamilton vector\cite%
{leach,goldstein2,sivardiere}, whose constancy is a direct consequence of
these ones of $\overrightarrow{L}$ and $\overrightarrow{\mathcal{A}}_{K}$.
Any linear combination of this two orthogonal vectors being conserved, we
can build a conserved vector corresponding to any vector of the plane. $%
\overrightarrow{\mathcal{S}}_{K}$ and $\overrightarrow{\mathcal{A}}_{K}$ are
particularly interesting choices because they give the directions of the
minor and major axes of the elliptical trajectory.

If we want that the functional forms of $a(r)$ and $b(r)$ being globally
defined during all the motion, we have to keep the same values for $\gamma $
and $\phi $ in every phase. With the choice $\phi =\frac{\pi }{2}$, Eq.(\ref%
{fradkinvect3}) becomes:

\begin{equation}
\mathcal{A}_{2n}=\mathcal{A}_{2n+1}=\gamma e^{2ni\Phi }  \label{fradkinvect4}
\end{equation}

$\overrightarrow{\mathcal{A}}$ is then constant when we pass from a even
phase to the following odd one, that is when we cross the apocenter. But
after a complete period of oscillation, when we reach the pericenter again,
we pass from $k=2n$ to $k=2n+1$. At this moment the value of $\mathcal{A}%
_{2n+1}$ change to$\mathcal{A}_{2n+1}$ with:

\begin{equation}
\mathcal{A}_{2n+2}=\mathcal{A}_{2n+1}e^{2i\Phi }  \label{rotation}
\end{equation}

This corresponds to a $2\Phi $ rotation of the associated vector. We recover
here the discontinuity jumps (observed by\ Serebrennikov,\ Shabad, Buch and\
Denman\cite{sereb,buch} and Peres\cite{peres} and first studied in a
detailed by Holas and March\cite{holas}) which make FBRS vector only a
piecewise conserved quantity: the perihelion vector $\mathcal{A}$ presents
discontinuities at each pericenter (with the choice made here and at each
apocenter if we choose the initial condition $r(t=0)=r_{M}$) corresponding
to a rotation of two times the apsidal angle.

\subsubsection{Harmonic oscillator, Kepler problem and Bertrand's theorem}

For $\Phi =\pi $, which is the case of the Kepler problem $U(r)=-r^{-1}$ for
every values of the characteristic parameters of the motion $L$ and $E$, $%
\overrightarrow{\mathcal{A}}$ is a true vector conserved quantity and is
identical to the usual Laplace-Runge-Lenz vector $\overrightarrow{\mathcal{A}%
}_{K}$. Indeed, in this case, with $E<0$, introducing Clairaut's variable%
\cite{whittaker} $u=\frac{1}{r}$, we have (see Eq.(\ref{g}) and(see Eq.(\ref%
{deff})):

\begin{equation}
g(r)=-\frac{L}{\sqrt{2}}\int_{u_{m}=\frac{1}{r_{m}}}^{u}\frac{dv}{\sqrt{%
-\left\vert E\right\vert +v-\frac{L^{2}}{2}v^{2}}}
\end{equation}

which gives\cite{grad} for $1-2L^{2}\left\vert E\right\vert =e^{2}>0$\cite%
{cordani}:

\begin{equation}
g(r)=\left[ \arg \sin \left( \frac{1-L^{2}v}{e}\right) \right] _{u_{m}}^{u=%
\frac{1}{r}}
\end{equation}

Since $r_{m}$ is the smallest root of $f(r)=\frac{2r^{2}}{L^{2}}\left(
-\left\vert E\right\vert +\frac{1}{r}-\frac{L^{2}}{2r^{2}}\right) =0$, that
is, $u_{m}=\frac{1}{r_{m}}$ is the greatest root of $-\left\vert
E\right\vert +u-\frac{L^{2}}{2}u^{2}=0$, we obtain:

\begin{equation}
u_{m}=\frac{1+e}{L^{2}}
\end{equation}

and:

\begin{equation}
g(r)=\arg \sin \left( \frac{1-\frac{L^{2}}{r}}{e}\right) +\frac{\pi }{2}
\end{equation}

Then (see Eq.(\ref{sol2}) with $\phi =\frac{\pi }{2}$):

\begin{equation}
\left\{ 
\begin{array}{c}
a(r)=\frac{\gamma }{r\sqrt{f(r)}}\cos \left( g\left( r\right) +\frac{\pi }{2}%
\right) =-\frac{\gamma }{r\sqrt{f(r)}}\sqrt{1-\left( \frac{r-L^{2}}{er}%
\right) ^{2}}=-\gamma \frac{L^{2}}{er^{2}} \\ 
b(r)=a(r)+\frac{\gamma }{r}\sin \left( g\left( r\right) +\frac{\pi }{2}%
\right) =-\gamma \frac{L^{2}}{er^{2}}-\frac{\gamma }{r}\frac{r-L^{2}}{er}%
=-\gamma \frac{1}{er}%
\end{array}%
\right.  \label{abkepler}
\end{equation}

Taking $\gamma =-e$, we recover the coefficients of the usual
Laplace-Runge-Lenz vector (see Eq.(\ref{LRL}):

\begin{equation}
\left\{ 
\begin{array}{c}
a(r)=\frac{L^{2}}{r^{2}} \\ 
b(r)=\frac{1}{r}%
\end{array}%
\right.
\end{equation}

In the isotropic harmonic oscillator problem (Hooke's problem) $\Phi =\frac{%
\pi }{2}$, for every $L$ and $E$. The global direction is then conserved but
the sense of $\overrightarrow{\mathcal{A}}$ is alterned at each pericenter
crossing : $\overrightarrow{\mathcal{A}}\rightarrow -\overrightarrow{%
\mathcal{A}}$\cite{sereb,buch}. The generalized Hamilton vector $%
\overrightarrow{\mathcal{S}}=\overrightarrow{L}\times \overrightarrow{%
\mathcal{A}}$ is subject to the same phenomenon : $\overrightarrow{\mathcal{S%
}}\rightarrow -\overrightarrow{\mathcal{S}}$. As for the Fradkin tensor $%
\mathcal{T}=\frac{1}{2}\overset{.}{\overrightarrow{r}}\otimes \overset{.}{%
\overrightarrow{r}}+\frac{\omega ^{2}}{2}\overrightarrow{r}\otimes 
\overrightarrow{r}$ , recalling that it can be written as\cite{sivardiere} $%
\mathcal{T}=\frac{\omega ^{2}}{2}\overrightarrow{\mathcal{A}}\otimes 
\overrightarrow{\mathcal{A}}+\frac{1}{2\left\vert \mathcal{A}\right\vert ^{2}%
}\overrightarrow{\mathcal{S}}\otimes \overrightarrow{\mathcal{S}}$ , we see
immediately that it's a global invariant of the motion.

As established by Bertrand\cite{bert,grandati} more than one century ago,
Kepler and Hooke potentials are the only central potentials for which the
apsidal angle is commensurable with $\pi $ for every values of the initial
parameters of the motion. For all the other central potentials, this
condition, which is necessary for the closure of the orbit, is obtained only
for specific values of $E$ and $L$. In these cases, as established by Holas
and March\cite{holas}, it is still possible for an orbit of multiplicity $n$
to build global geometrical invariants in form of $n$-arm stars by using the 
$n$ distinct FBRS vectors associated to the system. Nevertheless, as we have
seen before, the existence of a general (that is for every initial
condition) true invariant vector or tensor is a specificity of Kepler and
Hooke problems respectively.

\end{document}